\documentclass[review]{elsarticle}

\usepackage{hyperref}
\usepackage{amsmath,amssymb}
\usepackage{graphicx} 
\usepackage{dcolumn}  
\usepackage{bm}       

\usepackage[displaymath]{lineno}

\journal{Journal of Phys. Lett. B Templates}

\begin{document}

\begin{frontmatter}

\title{Pion Valence Quark Distributions from Maximum Entropy Method}

\author[IMP-CAS,UCAS]{Chengdong Han}

\author[IMP-CAS,SU-Chongqing]{Hanyang Xing}

\author[IMP-CAS,UCAS]{Xiaopeng Wang}

\author[IMP-CAS,UCAS,LZU]{Qiang Fu}

\author[IMP-CAS,IPNOrsay]{Rong Wang}
\cortext[mycorrespondingauthor]{Corresponding author}
\ead{rwang@impcas.ac.cn,wangrong@ipno.in2p3.fr}

\author[IMP-CAS]{Xurong Chen}
\cortext[mycorrespondingauthor]{Corresponding author}
\ead{xchen@impcas.ac.cn}

\address[IMP-CAS]{Institute of Modern Physics,Chinese Academy of Sciences, Lanzhou 730000, China}
\address[UCAS]{University of Chinese Academy of Sciences, Beijing 100049, China}
\address[SU-Chongqing]{School of Physical Science and Technology, Southwest University, Chongqing 400715, China}
\address[LZU]{Lanzhou University, Lanzhou 730000, China}
\address[IPNOrsay]{Institut de Physique Nucl\'eaire, CNRS-IN2P3, Univ. Paris-Sud,
Universit\'e Paris-Saclay, 91406 Orsay Cedex, France}

\begin{abstract}
Valence quark distributions of pion at very low resolution scale $Q^{2}_0 \sim 0.1$ GeV$^2$ are deduced from a maximum entropy method,
under the assumption that pion consists of only a valence quark and a valence anti-quark at such a low scale.
Taking the obtained initial quark distributions as the nonperturbative input in the modified Dokshitzer-Gribov-Lipatov-Altarelli-Parisi
(with the GLR-MQ-ZRS corrections) evolution, the generated valence quark distribution functions at high $Q^2$
are consistent with the measured ones from a Drell-Yan experiment. The maximum entropy method is also applied to estimate the valence quark
distributions at relatively higher $Q^2$ = 0.26 GeV$^{2}$. At this higher scale, other components (sea quarks and gluons)
should be considered in order to match the experimental data. The first three moments of pion quark distributions at high $Q^2$
are calculated and compared with the other theoretical predictions.
\end{abstract}


\end{frontmatter}


\section{Introduction}
\label{sec:intro}

At high energy, the scattering process with a hadron happens
on its internal constituents, namely the quarks and the gluons,
which is commonly called the partons. Parton distribution function (PDF) is the number
density information of the partons carrying some fractions of the hadron momentum.
Thanks to the factorization theorem based on quantum chromodynamics (QCD) theory,
the cross-section of high energy scattering on hadron is the product of the interaction
connecting the probe and the PDFs of the hadrons.
Therefore determination of PDFs of hadrons is an important project continually
in hadron physics study. Up to date, PDFs of nucleons are precisely determined
from the global analysis of worldwide experimental data. However, less
is know for the other hadrons, such as the pion. Pion is the lightest hadron,
acting as the key interaction carrier between nucleons, which attracts
a lot of interests from both experimentalists and theorists.

In experiment, the parton structure of pion is usually measured with
the muon pair production process of $\pi-N$ scattering\cite{Pion-NA3-data,Exp-E615},
the leading neutron production \cite{Pion-ZEUS-data,H1-ZEUS-data,Pion-H1-data}
$ep\rightarrow e'nX$ of deep inelastic scattering assuming the pion exchange is dominant,
and the prompt photon production of $\pi-N$ scattering \cite{Pion-WA70-data}.
The Drell-Yan data of $\pi-N$ scattering\cite{Pion-NA3-data,Exp-E615} accesses the valence
quark distribution of the projectile pion, while the leading neutron production
at HERA \cite{Pion-ZEUS-data,H1-ZEUS-data,Pion-H1-data} probes the sea component of pion
at small $x$, and the prompt photon production mainly constraints
the gluon distribution \cite{Aurenche1989}. The global fits of the pion PDFs
to the experimental data are performed by several groups, using
the next-to-leading-order QCD analysis \cite{SMRS1992},
using the constituent quark model parametrization \cite{GRS1998,GRS1999},
and using the modified DGLAP equations \cite{Lou2015}.

In theory, it is not simple to predict the distribution of the valence content,
since the pion is not only formed as a quark-antiquark system but also as one of
the Glodstone bosons in the chiral symmetry breaking of $SU(3)$ flavor.
However there are lots of progresses from the Dyson-Schwinger equations
(DSE) \cite{PionDS-2001,PionDS-2011,PionDS-2015,PionDS-2016,PionDS-2018},
the NJL model \cite{PionNJL,Davidson1995,Davidson2002}, light-front holographic QCD (LFHQCD) \cite{HLFHS-Pion},
Lattice QCD \cite{Best1997-LQCD},
chiral quark model \cite{PionChiral1998,Nam2012,PionChiral2016,Watanabe2018,PionChiral2018},
constituent quark model \cite{Szczepaniak1994,Frederico1994}
and QCD sum rule \cite{PionSumRule}, etc. Beyond the quark distribution functions,
the parton distribution amplitude and the generalized parton distributions of pion
are studied \cite{Arriola-PDA,Cloet-PDA,Broniowski-GPD-2008,Chang-GPD-2015,Mezrag-GPD-2015,Chouika-2017}.
The rainbow-ladder truncation of the DSE
well incorporate the dressed quark propagators of meson amplitudes
and the dynamical chiral symmetry breaking, giving a description
of the pion as a Goldstone boson. The discrepancy on the large-$x$ behavior
between DSE and the Drell-Yan data is thought to be the soft gluon resummation effect.
Currently, the generalized parton distributions of nucleon and pion based on LFHQCD
are determined with physical constraints. The analytic structure of pion valence
quark distribution in LFHQCD describe well the large-$x$ behavior of E615 data \cite{HLFHS-Pion}.
Lattice QCD so far only gives the reliable lower moments of PDFs.
Modeling the nonperturbative information of pion is a very challenging task,
which still needs a lot of efforts and more inventions.

In this work, we try to calculate the pion valence quark distribution
in a way as simple as possible. The simplest valence quark distributions
of pion at extremely low resolution scale ($\sim 0.1$ GeV$^2$) is provided
under the maximum entropy method (MEM) with the quark model constraints.
With the application of DGLAP equations with nonlinear corrections,
the valence quark distribution at high $Q^2$ generated from the simple nonperturbative
input is roughly consistent with the experimental measurements.
In Sec. \ref{sec:input}, the way to model the pion valence quark distributions
is shown. The definition of the entropy is shown in Sec. \ref{sec:MEM}.
The application of the modified DGLAP evolution is discussed in Sec. \ref{sec:ModDGLAP}.
The MEM results are compared to the experimental data and other models
in Sec. \ref{sec:results}. Some discussions and summary are given in Sec. \ref{sec:discussions}.

\section{A nonperturbative input for pion based on the quark model}
\label{sec:input}

The simplest description of the internal structure of hadrons is the quark model,
in which the meson consists of a quark-antiquark pair, while the baryon consists of three quarks.
This naive picture in quark model is usually confusing confronting the complicated
hadron structure observed in high energy scattering experiments.
The hadron structures display more than a quark-antiquark pair or three quarks in experiments.
The reason is that the resolving power of the probe is high. However,
assuming with a very low resolution probe (close to 0.1 GeV$^2$), the constituents
which can be resolved inside hadrons are merely the valence (constituent) ones, in the quark-parton model.

According to the quark model, the naive nonperturbative input of pion is composed of
a quark distribution and an anti-quark distribution. Taking $\pi^{+}$ as an example,
a reasonable hypothesis is that the pion consists of a up valence quark ($u_v$) and an
anti-down valence quark ($\bar{d}_v$) at extremely low resolution scale $Q_0^2$.
In this work, valence quark distribution functions at $Q_0^2$ are parameterized
to mimic the analytical solution of nonperturbative QCD.
The simplest function form to parameterize valence quark distribution is
the time-honored canonical parametrization $f(x) = A x^B (1-x)^C$, which describes
well the Regge behavior at small $x$ and counting rule at large $x$.
Therefore the parametrization of the naive nonperturbative input of $\pi^{+}$ is as follows:
\begin{equation}
u_v(x,Q_0^2)= \bar{d}_v(x,Q_0^2)=A_\pi x^{B_\pi}(1-x)^{C_\pi}.
\label{Parametrization}
\end{equation}
Here, the distributions of valence up quark and valence anti-down quark are
the same, assuming no breaking of isospin symmetry and that the mass difference
between up and down quark is trivial.

In quark model, a $\pi^+$ contains one up valence quark and one anti-down quark.
Therefore, we have the valence sum rule for the naive nonperturbative input as
\begin{equation}
\int_0^1 u_v(x,Q_0^2)dx= \int_0^1 \bar{d}_v(x,Q_0^2)dx=1.
\label{ValenceSum}
\end{equation}
For the naive nonperturbative input, the momentum sum rule at $Q_0^2$ is written as,
\begin{equation}
\int_0^1 x[u_v(x,Q_0^2)+\bar{d}_v(x,Q_0^2)]dx=1.
\label{MomentumSum}
\end{equation}
We assume that there are none of sea quarks and gluons at extremely low resolution scale $Q_0^2$.
The sea quarks and gluons at high $Q^2>Q_0^2$ are radioactively generated by the valence quark radiations.
If one wants to estimate the valence quark distribution at higher $Q^2$,
the momentum sum rule is modified as following,
\begin{equation}
\int_0^1 x[u_v(x,Q^2)+\bar{d}_v(x,Q^2)]dx=1 - g,
\label{MomentumSum2}
\end{equation}
in which $g$ is the fraction of the momentum carried by the components other than
the valence quarks, such as the sea quarks and gluons.

\section{Determination of pion valance quark distributions from maximum entropy method}
\label{sec:MEM}

With the constraints in Eq. (\ref{ValenceSum}) and (\ref{MomentumSum}),
only one free parameter is left for the nonperturbative input in Eq. (\ref{Parametrization}).
To determine the free parameter, the maximum entropy method is used.
According to the definition of the generalized information entropy of quarks in Ref. \cite{MEMPDF},
the entropy of valence quark distributions in pion is calculated as,
\begin{equation}
\begin{aligned}
S=&-\int_0^1 [ u_v(x,Q^2)\text{Ln}(u_v(x,Q^2))\\
  &+ \bar{d}_v(x,Q^2)\text{Ln}(\bar{d}_v(x,Q^2))] dx.
\end{aligned}
\label{EntropyFor}
\end{equation}
The most reasonable valence quark distributions are the ones when the entropy $S$ takes the maximum.
Note that, here the entropy is the generalized information entropy, and the natural logarithm
is used during the calculation. If the base of the logarithm changes, then the value of entropy changes.
The entropy difference matters instead of the absolute value. Therefore it does not matter
if the general entropy here is negative.

The definition in Eq. (\ref{EntropyFor}) is related to the Von Neumann entropy for the quantum object.
For a quantum object in the non-pure state, the Von Neumann entropy is defined as,
\begin{equation}
\begin{aligned}
S=-\text{Tr}\rho\text{Ln}(\rho),
\end{aligned}
\label{VonNeumannEntropyDef}
\end{equation}
in which $\rho$ is the density matrix in Hilbert space. Obviously, the valence quark
and the valence anti-quark at $Q_0^2$ are not in a pure quantum state due to the
strong interaction among them. In the collinear framework, parton distribution functions
contain the information describing the state of quarks. PDF $f_i(x)$ can be treated
as the density matrix $\rho$. Therefore Von Neumann entropy of the partons inside hadron
is written as,
\begin{equation}
\begin{aligned}
S=-\sum_i \int_0^1 f_i(x) \text{Ln}(f_i(x))dx,
\end{aligned}
\label{VonNeumannEntropyHadron}
\end{equation}
The only modification is that the trace operator becomes the integral operator over $x$.
The maximum entropy method in this work is actually applying maximum quantum entropy principle.

\section{Modified DGLAP evolution}
\label{sec:ModDGLAP}

With the maximum entropy method, the valence quark distributions at the low resolution scale
($Q^2<1$ GeV$^2$) are determined. However this nonperturbative parton information can not be compared to
the experimental measurements at the low $Q^2$ due to the parton-hadron duality.
At low $Q^2$, higher twist corrections or the hadron contribution to the scattering process
can not be ignored. To compare the nonperturbative input obtained from MEM to
the experimental measurement at high $Q^2$, we need a tool to evaluate $Q^2$-dependence
of parton distribution functions.

Dokshitzer-Gribov-Lipatov-Altarelli-Parisi (DGLAP) equations \cite{DGLAP-D,DGLAP-GL,DGLAP-AP}
are usually used to predict the parton distribution evolution over $Q^2$ scale.
However, in the kinematic region where the parton density is very high or the size
of parton is very big (low $Q^2$), parton-parton recombination correction can not
be neglected in the DGLAP evolution. The theoretical prediction of parton-parton recombination
process in addition to the splitting process is initiated by Gribov, Levin and Ryskin (GLR)\cite{GLR},
and followed by Mueller, Qiu (MQ)\cite{MQ}, Zhu, Ruan and Shen (ZRS)\cite{Zhu-1,Zhu-2,Zhu-3}
with concrete and different methods. In this work, the modified DGLAP evolution equations
with GLR-MQ-ZRS corrections are used to predict the scale-dependence of PDFs.
The modified DGLAP evolution based on the perturbative QCD calculation has been applied
to study the nucleon structure \cite{chen2014-protonPDF,AsySeaFit,IMParton-paper}
and the nucleon structure in nuclear medium \cite{Chen2014-EMC,nIMParton-paper}.
In this work, the DGLAP equations with the simplified GLR-MR-ZRS corrections are used,
which only the dominant gluon-gluon recombination process is included in the scale evolution \cite{IMParton-paper}.
DGLAP equations with gluon-gluon recombination corrections are written as,
\begin{equation}
\begin{aligned}
Q^2\frac{dxf_{q_i}(x,Q^2)}{dQ^2}
=\frac{\alpha_s(Q^2)}{2\pi}P_{qq}\otimes f_{q_i},
\end{aligned}
\label{ZRS-NS}
\end{equation}
for the valence quark distributions,
\begin{equation}
\begin{aligned}
Q^2\frac{dxf_{\bar{q}_i}(x,Q^2)}{dQ^2}
=\frac{\alpha_s(Q^2)}{2\pi}[P_{qq}\otimes f_{\bar{q}_i}+P_{qg}\otimes f_g]\\
-\frac{\alpha_s^2(Q^2)}{4\pi R^2Q^2}\int_x^{1/2} \frac{dy}{y}xP_{gg\to \bar{q}}(x,y)[yf_g(y,Q^2)]^2\\
+\frac{\alpha_s^2(Q^2)}{4\pi R^2Q^2}\int_{x/2}^{x}\frac{dy}{y}xP_{gg\to \bar{q}}(x,y)[yf_g(y,Q^2)]^2,
\end{aligned}
\label{ZRS-S}
\end{equation}
for the sea quark distributions, and
\begin{equation}
\begin{aligned}
Q^2\frac{dxf_{g}(x,Q^2)}{dQ^2}
=\frac{\alpha_s(Q^2)}{2\pi}[P_{gq}\otimes \Sigma+P_{gg}\otimes f_g]\\
-\frac{\alpha_s^2(Q^2)}{4\pi R^2Q^2}\int_x^{1/2} \frac{dy}{y}xP_{gg\to g}(x,y)[yf_g(y,Q^2)]^2\\
+\frac{\alpha_s^2(Q^2)}{4\pi R^2Q^2}\int_{x/2}^{x}\frac{dy}{y}xP_{gg\to g}(x,y)[yf_g(y,Q^2)]^2,
\end{aligned}
\label{ZRS-G}
\end{equation}
for the gluon distribution, in which the factor $1/(4\pi R^2)$ is from the normalization of
the two-parton densities, and $\Sigma$ is the sum of all quark distributions.
The parton splitting kernel functions $P_{qq}$, $P_{qg}$, $P_{gq}$,
$P_{gg}$ can be found in the literature \cite{DGLAP-AP}. The gluon-gluon recombination kernels
$P_{gg\to \bar{q}}$, $P_{gg\to g}$, and the two-parton density normalization $1/(4\pi R^2)$
can be found in the literature \cite{IMParton-paper}.

The running coupling constant of strong interaction is an important parameter for
the modified DGLAP evolution. Since the gluon-gluon recombination corrections are calculated
only in leading-order (LO) so far, the running coupling constant $\alpha_s$ at LO is used in the calculation.
The LO $\alpha_s$ is taken as the same one in Ref. \cite{IMParton-paper,GRV98}.

The initial hadron scale $Q_0^2$ of the nonperturbative input is also an important parameter to perform
the modified DGLAP evolution. The initial scale of the purely valence nonperturbative input
is at $Q_0^2 = 0.064$ GeV$^2$ \cite{MEMPDF} for the proton. In the rescaling model,
the initial resolution scale $Q_0^2$ depends on the size of the hadron.
Assuming that the initial scale $Q_0$ is anti-proportional to radius of hadron $R_{hadron}$ \cite{Q2-rescaling},
the initial scale $Q_0^2$ of the nonperturbative input for pion is estimated to be 0.108 GeV$^2$,
based on the world average data of charge radius \cite{PDG-HEP}. We found that the measurements
of pion charge radius from $\pi-e$ scattering and from $ep\rightarrow e\pi^+ n $ process
are quite different. If pion radius takes from the measurement of $ep\rightarrow e\pi^+ n $ process \cite{A1-Collab},
the initial scale $Q_0^2$ of the nonperturbative input for pion is estimated to be 0.0825 GeV$^2$.
In this work, the initial hadron scale $Q_0^2$ is rescaled due to the size of hadron.
Note that it is the $Q^2$ that is usually rescaled, instead of the $Q_0^2$, for the explanation
of the EMC effect \cite{Q2-rescaling}. ($Q^2$ is rescaled to a higher value due to the swelling of the nucleon
when embedded in nuclear medium.) Actually the method to rescale $Q_0^2$ or to rescale $Q^2$
leads to the same result, however the two rescaling factors are reciprocal to each other.

The initial scale of the nonperturbative input estimated in the paragraph above
is similar to the ones used in NJL model calculations \cite{PionNJL,Davidson1995,Davidson2002}.
For NJL model calculation in Ref. \cite{PionNJL}, the $Q_0^2$ is $0.18$ GeV$^2$ with $\Lambda_{QCD}=0.25$ GeV.
For NJL model calculations in Refs. \cite{Davidson1995,Davidson2002},
the $Q_0^2$ is $0.098$ GeV$^2$ with $\Lambda_{QCD}=0.226$ GeV, which is close to the square of the constituent quark mass.
For our estimation of the input scales of both proton and pion, the $\Lambda_{QCD}$ of the running
coupling constant is 0.204 GeV. Usually the smaller $\Lambda_{QCD}$ requires the smaller
input scale $Q_0^2$. The input scale of pion containing only valence quarks in this work is
small, and it is fixed by the input scale $Q_0^2$ of the proton, the proton radius and the pion radius.

\section{Results}
\label{sec:results}

\begin{figure}[htp]
\begin{center}
\includegraphics[width=0.6\textwidth]{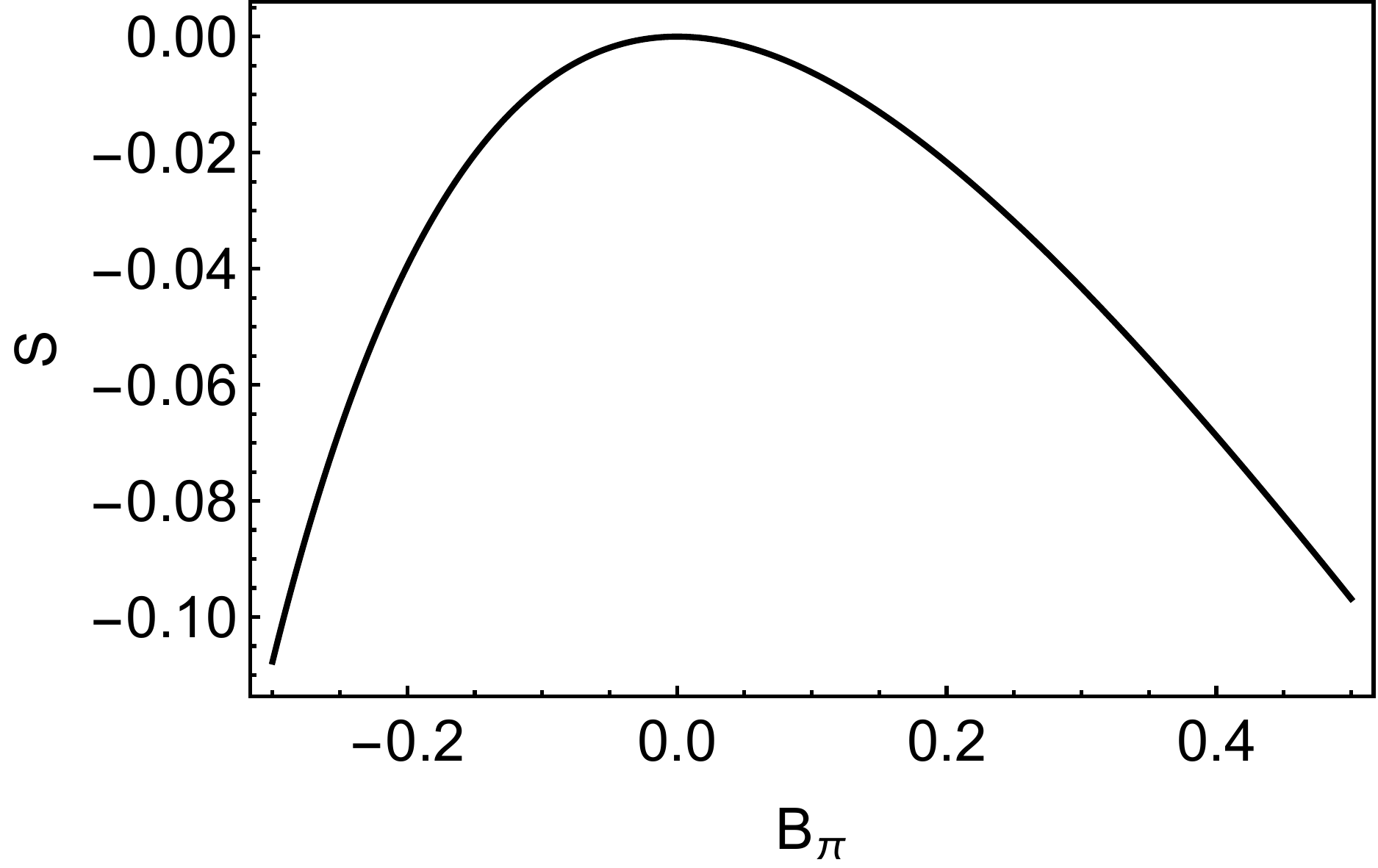}
\caption{
Information entropy $S$ of the purely valence quark nonperturbative input at $Q_0^2\sim 0.1$ GeV$^2$
is plotted as a function of $B_{\pi}$.
}
\label{fig:Entropy}
\end{center}
\end{figure}

\begin{figure}[htp]
\begin{center}
\includegraphics[width=0.6\textwidth]{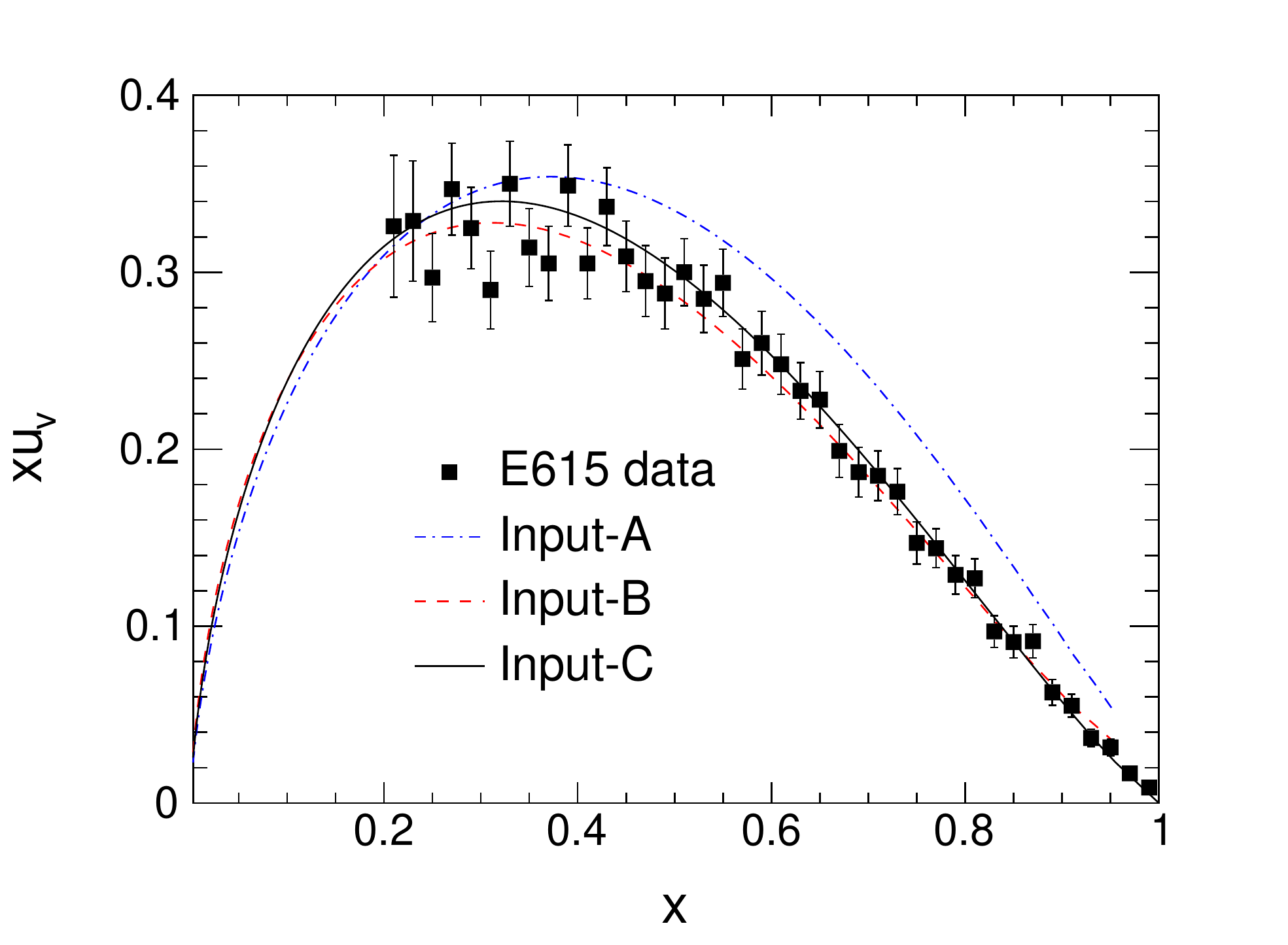}
\caption{(Color online)~
Comparisons among our predicted up valence quark momentum distributions with the experimental data
from E615 \cite{Exp-E615}, at $Q^2=20$ GeV$^2$. (See text for the explanations of Input-A, Input-B, and Input-C.)
}
\label{fig:E615}
\end{center}
\end{figure}

For the purely valence nonperturbative input, the distribution functions are obtained
to be $u_v(x, Q_0^2) = \bar{d}_v(x,Q_0^2) = 1$ where the entropy is at its maximum.
The entropy of the purely valence input as a function of $B_{\pi}$ is shown in Fig. \ref{fig:Entropy}.
It is very striking that the parton distribution is an uniform distribution, which is
the simplest function form to describe the internal structure.
It means that one component of the quark-antiquark pair can take any fraction of the hadron
momentum equality. By applying the modified DGLAP equations, the valence quark distribution
at high $Q^2$ from the two valence input is compared with the experimental data of E615 \cite{Exp-E615},
which is shown in Fig. \ref{fig:E615}.
Input-A is the initial valence quark distributions $u_v(x, Q_0^2) = \bar{d}_v(x,Q_0^2) = 1$
with $Q_0^2 = 0.108$ GeV$^2$; Input-B is the same initial valence quark distributions
but with $Q_0^2 = 0.0825$ GeV$^2$ (How we obtained these two starting scales $Q_0^2$ can be
found in the last two paragraphs in Sec. \ref{sec:ModDGLAP}).
Basically, the obtained purely valence nonperturbative inputs describe
the main feature of pion valence quark distribution. Especially, Input-B describes
the E615 data amazingly well.

\begin{figure}[htp]
\begin{center}
\includegraphics[width=0.6\textwidth]{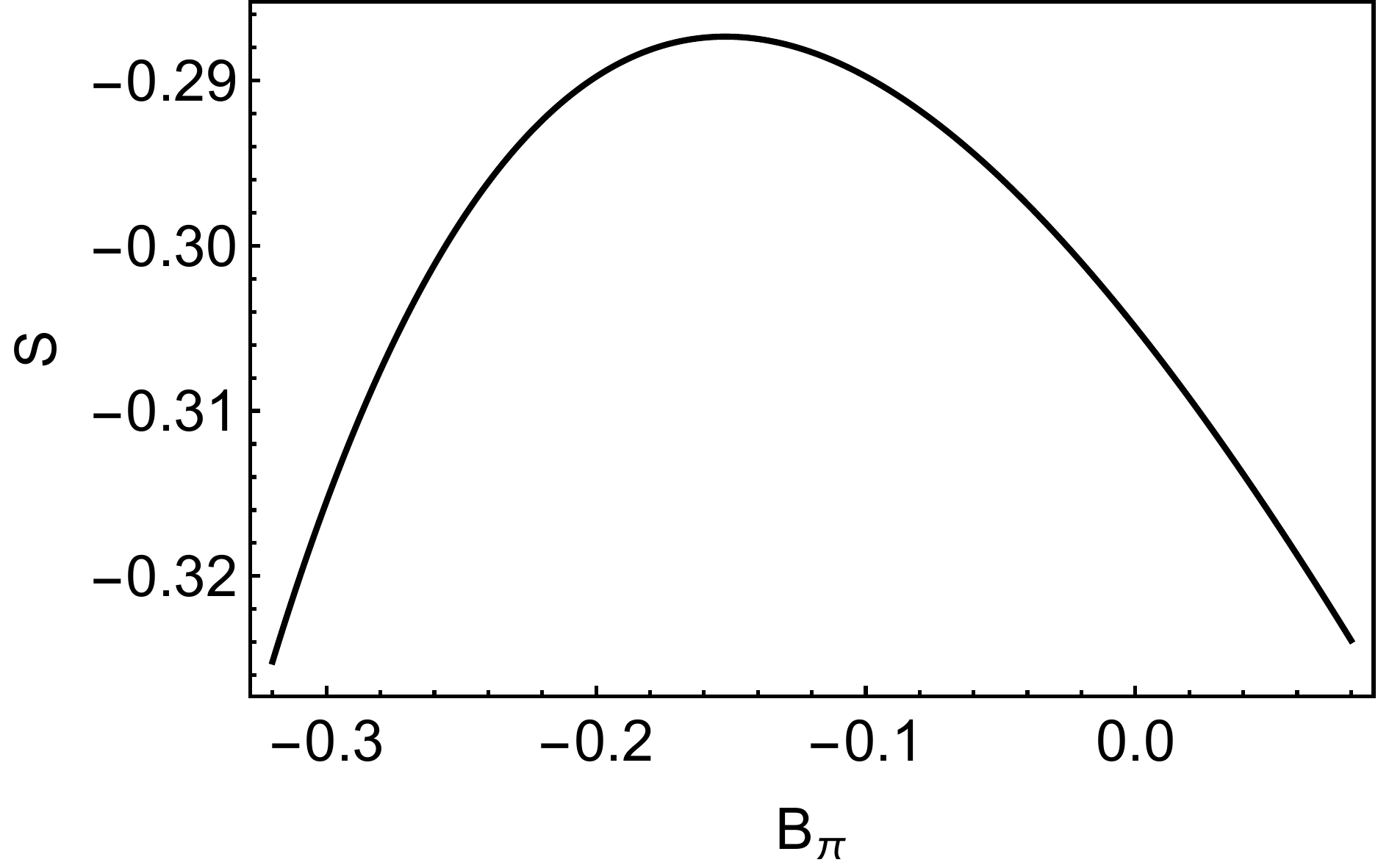}
\caption{
Information entropy $S$ of valence quarks at $Q^2=0.26$ GeV$^2$ is plotted as a function of the parameter $B_{\pi}$.
}
\label{fig:EntropyHigherQ2}
\end{center}
\end{figure}

In the global fit of pionic parton distributions \cite{GRS1998,GRS1999},
the valence quark distributions are usually parameterized at relatively high scale.
Taking GRS99 as an example, the valence quark distributions and the distributions
of other components are parameterized at $Q^2 = 0.26$ GeV$^2$.
An interesting test is that whether the maximum entropy principle can be
applied in this case or not. At the scale of $Q^2 = 0.26$ GeV$^2$, the momentum sum rule
is shown in Eq. \ref{MomentumSum2}, as the sea quarks and the gluons can not be ignored.
Taking $g=0.295$, we found that the valence quark distributions determined by the maximum entropy method
best match the E615 data. The valence quark distributions at $Q^2 = 0.26$ GeV$^2$
from MEM is given as $u_v = \bar{d}_v = 1.26x^{-0.152}(1-x)^{0.558}$, and it is labeled
as Input-C for the convenience of discussions.
The entropy under $g=0.295$ is shown in Fig. \ref{fig:EntropyHigherQ2} as a function of $B_\pi$.
The valence quark distribution at high $Q^2$ from Input-C is shown in Fig. \ref{fig:E615}.
The momentum fraction carried by the valence quarks of Input-C is consistent with the
calculation from a Dyson-Schwinger equation model \cite{PionDS-2001}.
In the Dyson-Schwinger equation model, valence quarks carry 71\% of the hadron momentum
at $Q^2 = 0.29$ GeV$^2$  \cite{PionDS-2001}. The momentum fraction carried by the valence quarks
at $Q^2 = 0.26$ GeV$^2$ are 75\% and 66\% from Input-A and Input-B respectively.
Therefore, Input-A, Input-B and Input-C are consistent with each other in terms of momentum fraction
carried by the valence quarks at $Q^2 = 0.26$ GeV$^2$.

Although the valence quark distributions are difficult to be calculated from first principle QCD,
the moments of valence quark distributions are calculated from Lattice QCD and some other models.
The moments of the momentum fraction of valence quark distribution are defined as following,
\begin{equation}
\begin{aligned}
<x^n>=\int_0^1 x^n u_{v}(x,Q^2) dx,
\end{aligned}
\label{MomenDef}
\end{equation}
with the superscript $n$ denotes the order of the moment.
The lowest three nontrivial moments of the valence quark distributions at $Q^2=4$ GeV$^2$
from the determined nonperturbative inputs in this work are listed in Table \ref{Tab:ccs},
compared to Lattice calculation, LFHQCD estimation, and some phenomenological model predictions.
The general agreements are found among them.

\begin{table}
\caption{
The list of the first three moments of valence quark momentum distributions from the QCD analyses,
the Dyson-Schwinger equations, the Lattice QCD, the LFHQCD, the chiral quark model
and our MEM estimations.
}
\label{Tab:ccs}
\begin{tabular}{c | c c c c }
\hline\hline
                    & $<x>_{u_v}^{\pi}$  & $<x^2>_{u_v}^{\pi}$  & $<x^3>_{u_v}^{\pi}$  & $Q^2$ (GeV$^2$) \\
\hline
QCD analysis \cite{SMRS1992}    & $0.23$        & $0.099$       &       $0.055$    &    4        \\
QCD analysis \cite{Wijesooriya2005}         & $0.217(11)$        & $0.087(5)$       & $0.045(3)$    &   27  \\
DSE \cite{PionDS-2016}                 & $0.26$        & $0.11$        & $0.052$    &   4  \\
LQCD \cite{Best1997-LQCD}                 & $0.273(12)$      & $0.107(35)$     & $0.048(20)$   &   5.76      \\
LQCD \cite{Detmold2003}                     & $0.24$        & $0.09$       & $0.043$    &   5.76  \\
LFHQCD \cite{HLFHS-Pion,TLiu-conversation}      & $0.233$        & $0.102$       & $0.056$    &   4  \\
Chiral quark \cite{Nam2012}         & $0.214$        & $0.087$        & $0.044$    &   27  \\
Chiral quark \cite{Watanabe2018}     & $0.23$        & $0.094$        & $0.048$    &   27  \\
NJL model \cite{Davidson1995,ArriolaViaEmail}        &  $0.236$  & $0.103$ & $0.057$ &  4  \\
This work, Input-A  & $0.27$           & $0.13$          & $0.074$     &   4        \\
This work, Input-B  & $0.24$           & $0.10$          & $0.058$     &  4     \\
This work, Input-C  & $0.24$           & $0.10$          & $0.057$     & 4        \\
\hline\hline
\end{tabular}
\end{table}

\section{Discussions and summary}
\label{sec:discussions}

For the lightest $q-\bar{q}$ confinement state, the maximum entropy method gives an extremely
simple form of the valence quark density distribution as the nonperturbative input,
which is an uniform distribution. It seems like that the momentum of the quark or the anti-quark
is totally uncertain. Therefore the spatial uncertainty of the valence quark in pion is
small, which makes us guess the color dipole is in a color transparency configuration \cite{Miller-CT-2011,Dutta-CT-2013}.
We notice that the naive nonperturbative distribution of no $x$-dependence
is as well the dynamical solution in NJL model with scalar and pseudoscalar
couplings \cite{PionNJL,Davidson1995,Davidson2002} and in the calculation of bound-state wave function
projecting on the null plane \cite{Frederico1992,Frederico1994}. Based on maximum entropy method,
this naive distribution is the simplest solution if we assume the function $Ax^B(1-x)^C$
to describe the nonperturbative input. What is the result
if we use more complex function to describe the nonperturbative input?
We have tried more complicated function forms,
such as $A x^B (1-x)^C(1+Dx)[1+E(1-x)]$ and $A x^B (1-x)^C(1+D\sqrt{x}+Ex)[1+F\sqrt{1-x}+G(1-x)]$.
The maximum entropy method gives $u_v(x,Q_0^2)=\bar{d}_v(x,Q_0^2)=1$ as well,
for these complex parametrizations. Therefore we argue that function form $Ax^B(1-x)^C$
is enough to describe the valence quark distributions at $Q_0^2$, and the uniform distribution
is the MEM solution. But why the simplest function $Ax^B(1-x)^C$ can be used to depict
the valence quark distributions should be explained with the dynamics from NJL model
or nonperturbative QCD.

With the modified DGLAP equations, the valence quark distributions at high $Q^2$
are calculated from the initial valence quark distributions deduced from the maximum entropy method.
Agreements are found between the MEM prediction and the Drell-Yan measurement by E615 Collaboration.
Maximum entropy method is also applied to study the valence quark distributions at $Q^2=0.26$ GeV$^2$.
All the quark distributions from MEM are basically consistent with each other, and they agree
with the E615 data. The moments of the valence quark distributions at $Q^2=4$ GeV$^2$ from MEM
are also compared to the Lattice QCD calculation and some other models, which shows
the consistency. The maximum entropy description of valence quark distribution is an alternative way in
understanding the main feature of pion structure. The small difference between MEM valence quark
distribution and the experimental measurement need to be addressed with QCD corrections.

The maximum entropy method can be used to get some information of hadron structure.
However, more constraints or dynamics of QCD theory should be used
to study the other type of structure such as the generalized parton distributions
and transverse momentum dependent parton distributions.
The calculations of the valence quark distributions are all done at very low
resolution scale with the maximum entropy method. This work also implies that DGLAP equations with
the twist-4 correction provide a good bridge connecting the nonperturbative information and
perturbative measurement.

The obtained valence quark distributions of pion can be further tested by the experiments
on electron ion colliders \cite{EICChina,US-EIC,CERN-EIC},
where the ``pion cloud" around the nuclei can be probed.

\noindent{\bf Acknowledgments}:
We thank Nu Xu for the enlightening and fruitful discussions.
This work is supported by the National Basic Research Program of China (973 Program) 2014CB845406.

\section*{References}


\end{document}